\documentclass[referee,useAMS,usenatbib, usedeluxetable]{mn2e}
\usepackage{graphicx}
\usepackage{epsfig}
\usepackage{float}
\usepackage{amsmath}
\usepackage{rotating}
\usepackage{natbib}
\usepackage{deluxetable}
\usepackage{lscape}
\usepackage[T1]{fontenc}
\usepackage[british]{babel}
\usepackage[varg]{txfonts}

\def \aj {AJ}
\def \apj {ApJ}

\def \apjl {ApJ}
\def \japa {JApA}
\def \nat {Nature}
\def \mnras {MNRAS}
\def \aap {A\&A}
\def \aaps {A\&AS}
\def \araa {ARA\&A}
\def \pasp {PASP}

\def \na   {New Astronomy}

\begin{document}
\title[INOV of AGN]
{Improved characterisation of intra-night optical variability of prominent AGN classes}
\author[Goyal et al.]
       {Arti Goyal,$^{1}$$\thanks{E-mail: arti.aries@gmail.com}$
       Gopal-Krishna,$^{1}$ Paul J. Wiita,$^{2}$
       C. S. Stalin$^{3}$ and Ram Sagar$^{4}$ \\
$^{1}$ Inter University Centre for Astronomy and Astrophyics (IUCAA), Post Bag 4 Ganeshkhind, Pune University Campus, Pune 411 007, India \\ 
$^{2}$ Department of Physics, The College of New Jersey, PO Box 7718, Ewing, NJ 08628-0718, USA. \\
$^{3}$ Indian Institute of Astrophysics (IIA), Bangalore 560 034, India. \\
$^{4}$ Aryabhatta Research Institute of Observational Sciences (ARIES), Manora Peak, Naini Tal 263 129, India. \\
}
\date{Received \today; accepted \today}
\pubyear{xxxx}
\volume{xx}
\date{Received xxx; accepted xxx}
\maketitle
\label{firstpage}
\begin{abstract}
The incidence of intra-night optical variability (INOV) is known to 
to differ significantly among different classes of powerful active
galactic nuclei (AGN). A number of statistical methods have been
employed in the literature for testing the presence of INOV in the 
light curves, sometimes leading to discordant results. In this paper we 
compare the INOV characteristics of six prominent classes of AGN, as 
evaluated using three commonly used statistical tests, namely the $\chi^2-$test, 
the modified $C-$test and the $F-$test, which has recently begun to gain popularity. 
The AGN classes considered are: radio-quiet quasars (RQQs),
radio-intermediate quasars (RIQs), lobe-dominated quasars (LDQs),
low optical polarization core-dominated quasars (LPCDQs),
high optical polarization core-dominated quasars (HPCDQs),
and TeV blazars. 
Our analysis is based on a large body of AGN monitoring data, 
involving 262 sessions of intra-night monitoring of a total 77 AGN, 
using 1-2 metre class optical telescopes located in India. In order 
to compare the usefulness of the statistical tests, we have also 
subjected them to a `sanity check' by comparing the number of false 
positives yielded by each test with the corresponding statistical 
prediction. The present analysis is intended to serve as a benchmark for 
future INOV studies of AGN of different classes.

\end{abstract}
\section{Introduction}
\label{intro}

Intensity variations at different wavebands 
offer a critical probe of the physics of 
active galactic nuclei (AGN), particularly because they remain unresolved 
at all wavebands. Intra-night optical variability (INOV),
or microvariability of an AGN is a phenomenon in which the 
optical flux density variations occur on hour-like to minute-like 
timescales, with amplitude ranging from a few 
hundredths to a few tenths of a magnitude. At optical wavebands, 
numerous such studies have been carried out, covering time 
scales down to hours and minutes, sometimes coordinated 
with monitoring in other wavebands (e.g., \citealt*{1989Natur.337..627Miller}; \citealt*{1995ARA&A..33..163Wagner}; 
\citealt{1995ApJ...452..582Jang,1997AJ....114..565Jang}; \citealt*{1999A&AS..135..477Romero}; 
\citealt{2002A&A...390..431Romero}; \citealt*{1993MNRAS.262..963GK, 1993A&A...271...89GK, 1995MNRAS.274..701GK}; 
\citealt{2000MNRAS.314..815GK,2003ApJ...586L..25GK,2011MNRAS.416..101GK};
\citealt*{1996MNRAS.281.1267Sagar}; \citealt{2004MNRAS.348..176Sagar};
\citealt{1991AJ....101.1196Carini, 1992AJ....104...15Carini, 2007AJ....133..303Carini}; 
\citealt*{1990AJ....100..347Carini, 1992ApJ...385..146Carini, 
1998AJ....116.2667Carini, 2003AJ....125.1811Carini, 2011AJ....141...49Carini}; 
\citealt{2004MNRAS.350..175Stalin, 2004JApA...25....1Stalin, 2005MNRAS.356..607Stalin};
\citealt{1997AJ....113.1995Noble};
\citealt{2007BASI...35..141AG, 2009MNRAS.399.1622AG, 2010MNRAS.401.2622AG, 2012A&A...544A..37AG};
\citealt*{2005A&A...440..855Gupta, 2009NewA...14...88Gupta};
\citealt{1998ApJ...501...69Diego}; \citealt{2009AJ....138..991Ramirez}; \citealt{2011MNRAS.412.2717Joshi};
\citealt{2008AJ....136.2359Gupta, 2008AJ....135.1384Gupta, 2012MNRAS.425.1357Gupta};
\citealt{2010ApJ...719L.153Rani, 2010MNRAS.404.1992Rani, 2011MNRAS.413.2157Rani};
\citealt{2010ApJ...718..279Gaur, 2012MNRAS.425.3002Gaur}).
One of the major finding of these studies, based on hundreds of nights 
of optical data, is that the INOV duty cycle (DC)
for the blazars and the radio-quiet quasars (RQQs)
differs significantly, being $\sim$60-70 
per cent for blazars and only $\sim$10 per cent for RQQs
(see \citealt{2002A&A...390..431Romero}; \citealt{2005MNRAS.356..607Stalin}; \citealt{2007AJ....133..303Carini} 
 and references therein). If one only
considers INOV with large amplitude 
(i.e., INOV amplitude $\psi$ $>$ 3 per cent), the DCs for blazars 
and RQQs are found to be $\sim$ 50 per cent and $\sim$0 per cent, 
respectively (\citealt{2003ApJ...586L..25GK}; \citealt{2005MNRAS.356..607Stalin}).     

According to the popular viewpoint, the origin of INOV in blazars 
is related to turbulence in the relativistic plasma jet, 
which gets amplified due to relativistic effects as the jet 
is directed close to our line of sight
(e.g., \citealt*{1995ARA&A..33..163Wagner}; \citealt{2003ApJ...586L..25GK}; also, \citealt*{1985MNRAS.215..383Singal}). 
In RQQs, however, INOV is thought to originate from the hot spots 
or flares on the accretion disk (\citealt*{1993ApJ...406..420Mangalam}; \citealt*{2006ASPC..350..183Wiita}) 
as they possess weak, poorly aligned jets (e.g.,\citealt*{1998MNRAS.299..165Blundell}; \citealt{2003ApJ...586L..25GK}). 
Hence, it becomes important to probe the INOV properties of different AGN 
classes, to develop understanding of the underlying physical processes.  
Since 1990, most observations of INOV have been made using CCD detectors, which allows
simultaneous recording of a number of stars within the same chip.
Not only are some of these simultaneously monitored stars used
for measuring any variations in the seeing disk during the course
of the monitoring session, but, more importantly, they are used
as non-varying standards relative to which the light curve of the
target AGN can be drawn. Thus, in differential photometry, 
the flux of the target AGN is divided by the flux of the 
comparison star recorded in the same CCD frame. 
The advantage of having the AGN and the comparison 
stars monitored in the same CCD frame is that they are subjected 
to essentially the same air mass and identical weather and instrumental conditions, 
and hence any variation in them are effectively cancelled out by taking the flux ratio.
Such `differential light curves' (DLCs) are 
also drawn for the candidate `comparison stars' as well, to check for 
the presence of INOV in any of those stars in which case such stars
are disqualified as comparison stars (e.g., 
\citealt{1991Sci...254R1238Miller}; \citealt{2004JApA...25....1Stalin, 
2006ASPC..350..183Wiita}).
The other advantage of using DLCs is that the effects of any
fluctuations in the atmospheric attenuation and even in the seeing
disk are mostly cancelled out, and this way the
variability detection threshold is pushed down enormously (e.g.,
\citealt*{1986PASP...98..802Howell}; \citealt*{1989Natur.337..627Miller}; \citealt{1993AJ....106.2441Gilliland};
 \citealt{2005PASP..117.1187Howell}). Thus, in our AGN sample, INOV with amplitudes as low as 1 to 2 per cent can be
routinely detected using 1-metre class telescopes.
(e.g., see \citealt{2012A&A...544A..37AG} and references therein). 
 
The most popular statistical test to detect the
variability in AGN DLCs has been the, so called, $C-$test, 
which was introduced by \citet{1997AJ....114..565Jang} and
\citet{1999A&AS..135..477Romero}.
Basically, this involves computation of a parameter `C' which  
is the ratio of the standard deviation of a given `target' AGN 
DLC to that of the comparison star-star DLC, i.e.,
\begin{equation}
C = \frac{\sigma_{t-s}}{\sigma_{s-s}} 
\end{equation}
\\
This ratio `C' has been taken to have a Gaussian distribution (e.g., 
\citealt*{1997AJ....114..565Jang}, \citealt{1999A&AS..135..477Romero}). 
Thus, an AGN whose DLC is found to have `C' greater than 
2.576 (the critical value corresponding to significance level, $\alpha=$ 0.01) 
is designated `variable'. Similarly, an AGN whose
DLC has `C' value greater than 1.950 but less than 2.576
(corresponding to $\alpha$ between 0.01 and 0.05) is termed as a `probable variable'.
However \citet{2010AJ....139.1269Diego} has 
questioned the validity
of this test on the ground that the $C$-statistic does not have
a Gaussian distribution and so the two tailed $p$-values should not be 
used as a statistical indicator of INOV at a given 
$\alpha$ (see also, \citealt{JApA2013AG}). 
According to him, the $C-$test is only valid provided the critical
values of $C$ are {\it properly} assigned at the chosen $\alpha$ (i.e., a modified $C-$test).

The ramification of the above is that the computation of 
INOV duty cycle for various AGN classes 
in the majority of the many studies cited above may not be reliable
as they are based on the $C-$statistic. The present 
work aims to address this anomaly by recomputing 
the INOV DCs for various AGN classes using ``proper'' 
statistical methods, for all the AGNs monitored under our 
AGN INOV programme at ARIES. This programme was launched 
in 1998 mainly using the 1-m Sampurnanand telescope 
(ST) of ARIES and has led to more than 250 
nights of intra-night monitoring of AGNs belonging to prominent AGN classes.
The present study aims
to recompute the INOV characteristics of these AGN classes, applying appropriate' 
statistical tests to these data.  
The other aim of 
this study is to examine the {\it robustness} of the statistical 
tests, by computing the number of {\it false positives}, 
or {\it Type 1 errors} for our large INOV data set. This will also 
serve as a {\it sanity check} on our analysis, a procedure 
generally lacking in INOV studies. 
Thus, we aim to compute INOV DC using three 
different methods ($\chi^2-, F-$ and modified $C-$test) 
to establish a uniform benchmark for future INOV studies.

\section{The Data Sample}
\label{sample}

A long-term programme for characterizing the 
INOV properties of prominent AGN classes was launched in 1998 , using the 1-m Sampurnanand telescope (ST) of ARIES. 
Usually, the
targets monitored in these studies are optically luminous 
and relatively bright point-like AGN, namely, quasars 
(both radio-loud and radio-quiet) and blazars.
Results of this ongoing study have been reported in a 
series of publications 
(\citealt{2012A&A...544A..37AG} and references 
therein; \citealt{2005MNRAS.356..607Stalin} and references therein). 
Optical intra-night monitoring data from other observatories 
in India, namely the 2-m Himalayan Chandra Telescope
(HCT) and the 2.4-m Vainu Bappu Telescope (VBT) of the Indian Institute of Astrophysics, 
the 1.2-m telescope of the Gurushikhar observatory of the Physical Research Laboratory 
and the 2-m Inter-University Centre for Astronomy and Astrophysics (IUCAA) Girawali Observatory (IGO) telescope 
were also obtained to augment the data taken 
with the 1-m ST. Nearly always, just one target AGN 
was monitored on a given night.  

The above intra-night monitoring program has covered 22 radio-quiet 
quasars (RQQs), 10 radio-intermediate quasars (RIQs), 
9 radio lobe-dominated quasars (LDQs), 23 radio core-dominated 
quasars including 11 showing high ($>$ 3 per cent) fractional optical polarization (HPCDQs), 
and 12 showing low ($<$ 2 per cent) fractional optical polarization (LPCDQs), as well 
as 13 BL Lac objects detected at GeV and/or TeV energies. Sources in these various 
classes were chosen from the catalog of 
\citet{2001yCat.7224....0Veron} and its subsequent releases.  
All of them lie at $z$ $> 0.14$ and have a listed 
$m_B < 18$mag, which allows adequate signal-to-noise 
ratio (SNR) in a typical exposure time of $\sim10$ minutes. 
Each source was monitored for a minimum duration of $\sim$4 hours. 
These CCD monitoring observations, followed by a careful and uniform
data analysis procedure, have routinely allowed INOV detection with 
amplitude ($\psi$) as low as 1 - 2 per cent. 
The present sample consists of 262 such intra-night observations 
derived from the entire data set of our ARIES AGN INOV programme.
The basic information on these sources 
 can be found in   
Table 1 which combines data from \citet{2004MNRAS.348..176Sagar}; \citet{2004MNRAS.350..175Stalin, 2004JApA...25....1Stalin, 2005MNRAS.356..607Stalin};
\citet{2007BASI...35..141AG, 2009MNRAS.399.1622AG, 2010MNRAS.401.2622AG, 2012A&A...544A..37AG};  and
\citet{2003ApJ...586L..25GK,2011MNRAS.416..101GK}.
\\
\section{Observations and data analysis}
\label{obs}
The observations were made mostly in the {\it R} filter and
occasionally in the {\it V} filter.
The exposure time was typically between 10 to 20 minutes for  
the ARIES and Gurushikhar observations but ranged between 
3 to 6 minutes for the observations using the VBT, IAO and IGO, 
depending on the brightness of the source, the phase of the 
moon and the sky transparency on that night. The field 
positioning was mildly adjusted so as to also include within 
the CCD frame at least 2--3 comparison stars. For all 
the telescopes, bias frames were taken intermittently, 
and twilight sky flats were also obtained. 

The pre-processing of the images (bias subtraction, flat-fielding 
and cosmic-ray removal) was done by applying the standard 
procedures in the 
{\textsc IRAF\footnote{\textsc {Image Reduction and Analysis Facility (http://iraf.noao.edu/) }} } 
and {\textsc MIDAS}\footnote{\textsc {Munich Image and Data Analysis System 
http://www.eso.org/sci/data-processing/software/esomidas// }}
software packages. The instrumental magnitudes 
of the target AGN (all point-like) and the stars in
the image frames were determined by aperture photometry, using
{\emph APPHOT}. 
The magnitude of the target AGN was measured relative to 
a few apparently steady comparison
stars present on the same CCD frame. In this way 
Differential Light Curves (DLCs) for each AGN were derived 
relative to 3 comparison stars designated as s1, s2, s3. 
Out of the resulting 3 star-star DLCs, we selected the steadiest star-star DLC 
(based on the lowest variance) for testing the INOV of the AGN 
monitored on that night. These two chosen stars were now labeled s1 and s2 and the 
corresponding AGN target DLCs were termed as `t-s1' 
and `t-s2' AGN DLCs and while `s1-s2' was the star-star DLC.

Mostly, the comparison stars used lie within about 1.5 magnitude
of the target AGN, this being an important criterion for minimizing 
the possibility of spurious INOV detection 
(e.g., \citealt{2007MNRAS.374..357Cellone}; \citealt{JApA2013AG}). 
Spurious variability on account of different second-order
extinction coefficients for the AGN and their comparison 
stars can be a potential problem if the optical colours of the objects 
are significantly different. Even though the {\it B-R} colors of the AGN 
and the comparison stars usually differ significantly in our 
sample, it was shown by \citet{1992AJ....104...15Carini}  
and \citet{2004JApA...25....1Stalin} that while
 the photons travelling through varying airmasses
during the course of monitoring will be differentially affected, this 
second-order term produces a negligible
effect on the derived INOV parameters.
For each night, an optimum aperture radius for photometry
was chosen by identifying the minimum dispersion
in the star-star DLCs, as a function of the chosen 
aperture radii, starting 
from the median seeing (FWHM) value on that night to 4 
times of that value. 
For a given night, we thus selected the aperture
size which yielded the minimum dispersion in the steadiest star-star DLC. 
This  also set the threshold for INOV detection on that night 
(see, \citealt{JApA2013AG}).
Typically, the selected aperture radius was 
$\sim$4$^{\prime\prime}$ and the effective seeing was $\sim$2$^{\prime\prime}$.

\section{Statistical methods for detecting variability}
\label{method}
In the present analysis, we have used 3 statistical tests 
($\chi^2-, F-$, and the modified $C-$ test) 
for the variability detection 
(see also, \citealt*{2010ApJ...723..737Villforth}).
Although it is argued by \citet{2010AJ....139.1269Diego} that the Analysis of Variance ($ANOVA$) 
statistic has the 
highest statistical power for detecting variability,
we could not use it here since  
many of our light curves had fewer than 30 data
points, precluding an effective application of the $ANOVA$ 
test.

\subsection{$\chi^2$-test of variance}
\label{chi2}
This test was first used by \citet{1983ApJ...272...11Pica} in the 
context of detecting long-term variability of 130 quasars and 
BL Lacs over a continuous monitoring duration of 13 years, to detect any
variations below the visual detection limit. Usually, 
in variability detection the {\it null hypothesis}
used is that a flat line fits the light curve and hence the object is non-variable.
The {\it null hypothesis} is rejected 
when the statistic exceeds a critical value for a given significance level, $\alpha$. It is given as 

\begin{equation}
\label{chi2test}
\chi^2 = \sum\limits_{i=1}^{N_p-1}\frac{(x_i - \bar{x})^2}{\sigma_i^2} ,
\end{equation}
where $N_p$ is the number of data points, $x_i$ is the 
magnitude of the $i^{th}$ data point in the lightcurve, 
$\bar{x}$ is the average over all $x_i$ and $\sigma_i$
is the rms measurement error associated with each $x_i$.  

\subsection{$F$-test of variance}
\label{Ftest}

The $F-test$, first used by \citet{1988AJ.....95..247Howell} 
to characterise  stellar variability, has recently been highlighted 
by \citet{2010AJ....139.1269Diego} in the context of INOV. 
The $F-$statistic compares the
observed to expected variances. The null hypothesis is 
rejected when the ratio exceeds a critical value for a chosen value of 
$\alpha$ (see, also, \citealt*{2010ApJ...723..737Villforth}). 
It is given as : 
 
\begin{equation}
\label{ftest}
F =\frac{Var_{observed}}{Var_{expected}}= 
\frac{Var_{t-s}}{\langle \sigma_{t-s}^2 \rangle}
\end{equation}
where $Var_{t-s}$ is the variance of the `target-star' DLC, and
$\langle \sigma_{t-s}^2 \rangle$,
is the mean of the squares of the (formal) rms errors of the individual data
points in the `target-star' DLC.

A  different form of $F-$test, the so called 
``scaled $F-$test'' (e.g., \citealt{2011MNRAS.412.2717Joshi}, 
see also, \citealt{1988AJ.....95..247Howell}) has 
been used to determine variability in AGN DLCs 
with an aim to compensate for the brightness mismatch 
between the target AGN and comparison stars used to 
derive the DLCs, which may lead to spurious variability 
in DLCs due to varying photon statistics. In this approach, 
to compensate for the brightness mismatch between the `target' AGN and the
comparison stars, a factor $\kappa$ is computed as

\begin{equation} 
\kappa = \frac{\sum\limits_{i=1}^{N_p} \sigma^2_{i,err} (t - s1)/N_p}{ \sum\limits_{i=1}^{N_p} \sigma^2_{i,err}(s1 - s2)/N_p}, 
\end{equation}
where $\sigma_{i,err} (t - s1) $ and $\sigma_{i,err}(s1 - s2)$
are, respectively the rms errors on $i^{th}$ data point of the 
target-star and star-star DLCs as returned by the {\sc APPHOT/IRAF} 
routine and $N_p$ is the number of data points in a given DLC.
In words,  $\kappa$ is the ratio of the mean of the ``expected'' variances between the target and and star 1, to the mean of the 
expected variances between the two stars, where ``expected" indicates the values are taken from IRAF photometery.
Then the `scaled' $F-$test (or $F_s$)  becomes
\begin{equation} 
F_s = {\frac{var(t - s1)}{ \kappa .  var(s1 - s2)}}, 
\end{equation}
where $var(t-s1)$ is the variance of the $t-s1$ DLC and $var(s1-s2)$ is the 
variance of the $s1-s2$ DLC.
Plugging in $\kappa$ from above gives
\begin{equation}
F_s = \frac{var(t-s1)}{var(s1-s2)} / \frac{mean(var(t-s1))}{mean(var(s1-s2))},
\end{equation}
so
\begin{equation}
F_s = \frac{var(t-s1)}{mean(var(t-s1))} / \frac{var(s1-s2)}{mean(var(s1-s2))}.
\end{equation}

Now, the $F$ for the target AGN using s1 as a control star is actually
$F(t,s1) = var(t-s1)/mean(var(t-s1))$. 
Accordingly, the $F$ for s1 using s2 as control is $F(s1,s2) = var(s1-s2)/mean(var(s1-s2)).$
So, what we call $F_s$ is:
$F_s = F(t,s1)/F(s1,s2)$.
The problem with this method is that $F_s$ does not follow a
$F-$distribution, since it is the ratio of F(AGN,s1) to F(s1,s2), as
shown above.  However, the ratio of two functions each distributed as $F$, is not distributed
as $F$ and hence one should not use the critical values associated with 
$F-$distribution at a given $\alpha$, as originally done in 
\citet{1988AJ.....95..247Howell} and \citet{2011MNRAS.412.2717Joshi}. 
Since we do not know the distribution of the 
ratio of F-distributions, the use of the `scaled' $F-$test 
in its present form is a poor choice. Instead, the standard $F-$test
as given in Equ. \ref{ftest} should be used; this adequately takes care of 
the brightness mismatch between the AGN and the comparison stars.

\subsection{Modified $C$-test of variance}
\label{Ctest}

The $C-$statistic is given as (see, e.g., \citealt*{2010ApJ...723..737Villforth}) 
\begin{equation}
\label{ctest}
C =\frac{\sigma_{observed}}{\sigma_{expected}}= 
\frac{\sigma_{t-s}}{\langle \sigma_{t-s} \rangle},
\end{equation}
where $\sigma_{t-s}$ is the standard deviation of the `target-star' DLC, and
$\langle \sigma_{t-s} \rangle$
is the mean of the (formal) rms errors of the individual data
points in the `target-star' DLC.

As already explained in Sect. \ref{intro}, it is improper to 
use {\it p-} values of a Gaussian distribution to characterise 
this statistic. 
For a given $\alpha$, the real critical values of the $C-$ statistic 
are equal to the square roots of the critical values for
the $F-$ statistic (\citealt{2010AJ....139.1269Diego}, 
see also, \citealt*{2010ApJ...723..737Villforth}). We term
this as the `modified $C-$test'.

Since all of the above three methods associate a flux/magnitude estimate 
with its error estimate, it is important to determine the
photometric errors accurately. As emphasized 
in several independent studies, the 
photometric errors returned by $APPHOT$ are 
significatnly underestimated (\citealt{2004JApA...25....1Stalin}; \citealt{JApA2013AG}
 and references therein). \citet{JApA2013AG}
have reported the latest attempt to determine this under-estimation 
factor, $\eta$, using an unprecedented data set consisting 
of 262 steady star-star DLCs. They find $\eta= 1.54\pm0.05$ which 
confirms the previous estimates by the same 
group which were based on much smaller data sets. \citet{JApA2013AG} also showed that the 
determination of $\eta$ is quite insensitive to the magnitude 
difference of upto 1.5 mag between the star pair used for deriving a DLC. 
Thus, $\eta=$1.54 has been used in the present analysis to
scale up the {\sc IRAF} photometric magnitude errors.

\section{Determination of INOV parameters}
\label{parameters}
We have used three different statistical tests, 
$\chi^2-$ $F-$ and the modified $C-$test, to determine
the presence of INOV in our target AGN DLCs (t-s1 and t-s2). 
To correct for the photometric error underestimation, 
as stated in previous section, 
the denominator of the test statistic has been replaced with
$\eta^2\sigma_i^2$ in case of the $\chi^2-$test, $\eta^2\sigma^2$ 
in case of the $F-$test and $\eta\sigma$ in case of the modified $C-$test,
before subjecting the AGN DLCs to the statistical tests.  
The significance level set for a given test determines
the {\it expected} number of {\it false positives} 
(i.e., the robustness of the test).
Since we aim to compare 
the variability results obtained using the three different methods, 
the comparison must be made at the same chosen significance level. 
Therefore, we have chosen two significance levels, 
$\alpha = $ 0.01 and 0.05, corresponding to $p-$values
of $\ga$ 0.99 and $\ga$ 0.95, respectively for 
each of the three tests. Recall that smaller the value 
of $\alpha$ is, the less likely it is to occur by chance. 
Thus, in order to claim a genuine 
INOV detection, i.e., assigning a `variable'
designation (V), we stipulate that the computed statistic 
value is above the critical value corresponding to $p > 0.99$ (i.e., $\alpha=$ 0.01)
for a given degree of freedom ($\nu = N_p - 1$).
We assign a `probable variable' (PV) designation when
the computed test statistic value is found to be between the
critical values at $\alpha = $ 0.01 and 0.05, otherwise
`non-variable' (N) designation as assigned to a DLC.

\subsection{ Peak-to-peak INOV amplitude ($\psi$)}

The peak-to-peak INOV amplitude is calculated using the definition of 
\citet{1999A&AS..135..477Romero}
\begin{equation} 
\psi= \sqrt{({D_{max}}-{D_{min}})^2-2\sigma^2} 
\end{equation}
with  $D_{min,max}$ = minimum (maximum) in the AGN differential light curve,
and $\sigma^2$= $\eta^2$$\langle\sigma^2_{i}\rangle$ where $\sigma_i$ 
is the error associated with each data point and $\eta$ =1.54 (Sect. \ref{method}).

\subsection{Computation of INOV duty cycle (DC)}
\label{inov_dc}
The INOV duty cycle was computed following the definition of
\citet{1999A&AS..135..477Romero} (see, also, \citealt{2004JApA...25....1Stalin} ):
\begin{equation} 
DC = 100\frac{\sum_{i=1}^n N_j(1/\Delta t_j)}{\sum_{j=1}^n (1/\Delta t_j)} \hspace*{0.5cm} {\rm per~cent} 
\label{eqno1} 
\end{equation}
where $\Delta t_j = \Delta t_{j,obs}(1+z)^{-1}$ is the duration of the
monitoring session of a source on the $j^{th}$ night, corrected for
its cosmological redshift, $z$. Note that since for a given source the
monitoring durations on different nights were not always equal, the
computation of DC has been weighted by the actual monitoring duration
$\Delta t_j$ on the $j^{th}$ night. $N_j$ was set equal  to 1 if INOV
was detected, otherwise $N_j$ = 0.

In order to compute the INOV DC for our samples of different AGN classes, 
we first assigned
the variability classification to each AGN observation by 
carrying out each test independently for both the AGN DLCs
(drawn relative to the comparison stars s1 and s2, Table \ref{result}). 
This has yielded two estimates 
of the INOV DC for each AGN class and the results are listed in Table \ref{dc_stat}. 
It is seen that for a few of our observations, there is a difference between 
the AGN INOV status
inferred using its two DLCs drawn relative to two comparison stars (Table \ref{result}). 
A possible explanation is that one of the stars may itself have varied.
Since any such putative low-level INOV of the comparison star would 
remain unnoticed by eye, we have no justification to prefer one 
comparison star over the other (in terms of steadiness of the DLC). 
We have therefore listed in Table \ref{dc_stat} the
estimates of INOV duty cycle (DC) for each AGN using both chosen comparison
stars, s1 and s2. Also, in cases where the `steady' star-star DLC 
is itself termed as `V' or `PV', we have ignored the variability
inferred for the corresponding AGN DLCs
while computing the final DC. We quote the final duty cycles of 
INOV for each  AGN class
by taking the average of the estimates using the two DCs
(i.e., using the two comparison stars).

Table \ref{dc_stat} summarizes our INOV DC estimates for 
the 6 AGN classes,  namely, RQQs, RIQs, LDQs, LPCDQs, HPCDQs and blazars.
 Note that in our sample of blazars, 
we have clubbed together the HPCDQs and 
BL Lac objects (\citealt*{1980ARA&A..18..321Angel}; \citealt{1992ApJ...398..454Wills}).
All of them are detected at GeV energies by the $FERMI$ 
(\citealt{2010ApJ...715..429Abdo}). Out of 22 blazars in our sample, 11 blazars 
(marked with an asterisk in the Table \ref{result}) are detected at 
TeV energies\footnote{http://tevcat.uchicago.edu/} and the remaining (11 blazars) 
have relatively harder $\gamma-$ray photon indices ranging between 2.02--2.63 
(\citealt{2010ApJ...715..429Abdo}). These blazars with harder GeV spectrum 
can be termed as candidate TeV blazars (\citealt{2013ApJ...764..119Senturk}). 
We have  computed the INOV DC of both confirmed TeV blazars 
and candidate TeV blazars in our current analysis and have called them all TeV blazars, for simplicity.
The INOV DC estimates using the $\chi^2-$test are: 
22 per cent for RQQs (25 per cent if PV cases are also included; 68 nights), 
30 per cent for RIQs (42 per cent if PV cases are also included; 31 nights),
19 per cent for LDQs (still 19 per cent if PV cases are also included; 35 nights), 
17 per cent for LPCDQs (26 per cent if PV cases are also included; 43 nights),
49 per cent for HPCDQs (55 per cent if PV cases are also included; 31 nights) 
and 43 per cent for TeV blazars (48 per cent if PV cases are also included; 85 nights).
If only INOV with $\psi > $ 3 per cent is considered, the INOV DCs 
are $\sim$8, $\sim$12, $\sim$6, $\sim$8, $\sim$38, $\sim$27 per cent
for RQQs, RIQs, LDQs, LPCDQs, HPCDQs, and TeV blazars, respectively.   

The INOV DCs using the $F-$test and the modified $C-$test are found to be 
indistinguishable. Their values of DCs are :
10 per cent for RQQs (23 per cent if PV cases are also included; 68 nights),  
18 per cent for RIQs (23 per cent if PV cases are also included; 31 nights),
5  per cent for LDQs (10 per cent if PV cases are also included; 35 nights), 
17 per cent for LPCDQs (28 per cent if PV cases are also included; 43 nights),
43 per cent for HPCDQs (51 per cent if PV cases are also included; 31 nights) 
and 45 per cent for TeV blazars (53 per cent if PV cases are also included; 85 nights).
If INOV with $\psi > $ 3 per cent is considered, the INOV DCs 
are $\sim$6, $\sim$11, $\sim$3, $\sim$10, $\sim$38, $\sim$32 per cent
for RQQs, RIQs, LDQs, LPCDQs, HPCDQs, and TeV blazars, respectively.

\section{Discussion}

An important step involved in the comparison presented here is 
to check the {\it robustness} of each statistical test. 
For this, we have also calculated the observed number of false positives 
({\it `Type 1 error'}) for each of the 3 tests. A  false positive
arises due the rejection of true null hypothesis by a test, when 
applied to the non-varying DLC (i.e., the inability 
to discern a non-variable object as non-variable). To compute the number of 
false positives, we have subjected the `steady' 
star-star DLCs to the 3 afore-mentioned statistical tests. Assuming
{\it a priori} that the star-star DLCs are steady, 
outcome of the statistical test applied to them 
should be consistent with the  expected number of 
 false positives for the assumed value of $\alpha$. 
We have adopted the same variability criterion for detecting INOV in the `steady' 
star-star DLC as that applied for the AGN DLCs (see Sect. \ref{parameters}). 
The number of  false positives depends only 
on the number of the star-star DLCs examined and the value of $\alpha$ chosen for 
the test. Thus, if the number of  false positives is found to be much
different from the  expected number, the test cannot be 
deemed reliable. 

We note that for our entire data set of 262 steady star-star DLCs, the {\it expected}
numbers of  false positives, at $\alpha=$ 0.01 and 0.05, are $\sim$ 3 and $\sim$13, 
respectively.
Since the distribution of false positives (Type 1 errors)
is binomial, we expect its actual number for a given test will be between 0
and 9 and in most cases between 3$\pm$2 at $\alpha = 0.01$. Similarly, at $\alpha = 0.05$,
the actual number of false positives is between 2 and 24 and
in most cases is 13$\pm$ 4.
The {\it observed} numbers of false positives reported by the 
application of the $\chi^2$ statistic for our data set are 43 
and 59, respectively, for $\alpha=$ 0.01 and 0.05.
Similarly, the numbers of  false positives reported by the 
application of the $F-$ statistic, at $\alpha=$ 0.01 and 0.05, 
are 6 and 18, respectively. Finally, the  observed numbers 
of false positives reported by the application of the modified 
$C-$ statistic, at $\alpha=$ 0.01 and 0.05, are 6 and 18, respectively.
We thus find that the {\it observed} and {\it expected} numbers of 
false positives are significantly different only when the
`steady' star-star DLCs are subjected to the $\chi^2$ statistical test. 
However, this is not so when the `steady' star-star DLCs are subjected 
to the $F-$test and the modified $C-$test, for which the {\it observed} 
and {\it expected} numbers of false positives
 are in close agreement. 
Thus, we conclude that the $F-$test and the
modified $C-$test can be used for variability detection, while
the $\chi^2-$test is not a reliable variability estimator.

\section{Summary}
\label{summary}
In this study, we have determined the INOV duty cycles for 
6 prominent AGN classes, namely RQQs, RIQs, LDQs, LPCDQs, HPCDQs 
and TeV blazars, using the data acquired under our 
AGN INOV programme at ARIES. 
The INOV DCs for these classes are found to be about : 
10 per cent for RQQs (68 nights), 18 per cent for RIQs (31 nights),
5 per cent for LDQs (35 nights), 17 per cent for LPCDQs (43 nights),
43 per cent for HPCDQs (31 nights) and 45 per cent for TeV detected blazars 
(including HPCDQs, 85 nights). 
These values are obtained using the results of the $F-$test
and the modified $C-$test, whereas the $\chi^2-$test is not found to be a reliable 
estimator for INOV as shown in the present study.

We note that these improved estimates of DCs are in fairly good agreement with 
the previous estimates (e.g., \citealt{2003ApJ...586L..25GK}; 
\citealt{2005MNRAS.356..607Stalin}; \citealt{2007AJ....133..303Carini}) based on the application 
of $C-$test introduced by \citet{1997AJ....114..565Jang}. However, unlike
the previous studies, we find that DCs for INOV amplitude 
$\psi >3$ per cent is not {\it zero} for RQQs, RIQs and LDQs, 
although it is still found to be much smaller than that for the blazar class 
(Table \ref{dc_stat}, see also, \citealt{2003ApJ...586L..25GK}; \citealt{2004MNRAS.350..175Stalin}; \citealt{2010MNRAS.401.2622AG}). 
This is understandable, as the $C-$test in its
original form  given by \citet{1997AJ....114..565Jang}, is more conservative
than the $F-$test, or the modified $C-$test 
(see \citealt{2010AJ....139.1269Diego}). Note that for $\psi>$ 3 per cent, the INOV DC for 
TeV blazars is $\sim$32 per cent, slightly less 
than that for the HPCDQ class ($\sim$ 38 per cent). However, this small difference
can be explained by the inclusion of
the well known BL Lac object PKS 0735+178 in the TeV blazar sample. 
This BL Lac object is known for its 
intriguing lack of INOV 
(see, \citealt{2009MNRAS.399.1622AG}). If we exclude 
PKS 0735+178 (i.e., 20 out of the 85 nights devoted to TeV blazars), we find that
the INOV DC for the TeV blazars increases to $\sim$ 40 per cent, 
which is close to the value found for HPCDQs, as indeed expected. 

In this study, we have also examined the robustness of the 3 tests by computing
the  observed numbers of  false positives for our entire sample of star-star DLCs. This has 
provided a sanity check on our analysis procedure.  We thus conclude
that the $\chi^2-$test is not a reliable estimator for INOV detection in AGN 
light curves. Instead, the $F-$test, or the modified $C-$test, should be used. 
INOV DCs estimated here using the refined statistical methods
should serve as the reference for future INOV studies of
prominent AGN classes.

\clearpage
\newpage

\begin{landscape}
\begin{deluxetable}{cccccccccccccccccccccc}
\tablecolumns{21}
\tabletypesize{\tiny}
\tablecaption{Summary of the INOV observations and the derived variability status for the target AGN and the corresponding `steady' star-star DLCs \label{result}}
\tablewidth{0pt}
\tablehead{
AGN  & Obs. date    & Tel.{$^\P$}  &  Fil.  & Dur. & $N_p$ & $\sigma_{s1-s2}$ & $\psi$  & $\Delta m_{t-s1}$ & $\Delta m_{t-s2}$   & $\Delta m_{s1-s2}$   &   \multicolumn{3}{|c|}{$\chi^2-test $:value (status$)^\dag$}   &  \multicolumn{3}{|c|} {$F-test$:value (status)$^\dag$ } & \multicolumn{3}{|c|} {modified $C-test$:value (status)$^\dag$ } &  Ref$^\pounds$.  \\
     &    dd.mm.yy  &              &        &  (hr)&       & $(\%)$ & $(\%)$  & (mag) & (mag)   & (mag)   & $t-s1$ & $t-s2$ & $s1-s2$  &   $t-s1$ & $t-s2$ & $s1-s2$  &  $t-s1$ & $t-s2$ & $s1-s2$  &   \\
        (1) & (2) & (3) & (4) & (5) & (6) & (7) & (8) & (9)  & (10) & (11) & (12) & (13) & (14) & (15) & (16) & (17) & (18) & (19)  & (20) & (21)  \\ }
\startdata
\multicolumn{21}{l}{\small Radio quiet quasars (RQQs)[22 sources; 68 DLCs]}\\ \hline	
&&&&&&&&&&&&&&&&&&&&\\	
J0045$+$0410 & 21.10.98 & ST  & R & 2.39 & 14 & 0.2 & 3.06 & -0.08 & -0.46 & -0.38 & 49.60(V)  & 33.77(V)  & 1.84(N)  & 2.91(PV) & 2.01(N)  & 0.13(N) & 1.70(PV) & 1.42 (N) & 0.36(N) & 	   (a) \\ 
J0045$+$0410 & 05.11.98 & ST  & R & 3.21 & 30 & 0.9 & 4.89 & -0.11 & -0.48 & -0.37 & 32.96(N)  & 25.85(N)  & 16.88(N) & 1.10(N)  & 0.72(N)  & 0.55(N) & 1.05(N)  & 0.85(N)  & 0.74(N) & 	   (a)\\  
J0045$+$0410 & 16.10.04 & HCT & R & 6.04 & 25 & 0.2 & 1.22 & -0.35 & 1.51  & 1.86  & 54.44(V)  & 68.30(V)  & 33.46(N) & 2.31(PV) & 2.84(V)  & 1.24(N) & 1.51(PV) & 1.68 (V) & 1.11(N) & 	   (b)\\
&&&&&&&&&&&&&&&&&&&&\\	
J0103$+$0321 & 05.11.05 & HCT & R & 5.94 & 21 & 0.4 & 1.73 & 1.40  & 0.31  & -1.09 & 11.59(N)  & 15.51(N)  & 15.31(N) & 0.60(N)  & 0.83(N)  & 0.72(N) & 0.77(N)  & 0.91(N)  & 0.85(N) & 	   (b)\\
J0103$+$0321 & 05.11.05 & ST  & R & 5.83 & 20 & 0.6 & 2.84 & 1.43  & 0.86  & -0.57 & 14.61(N)  & 17.70(N)  & 18.59(N) & 0.66(N)  & 0.83(N)  & 1.02(N) & 0.81(N)  & 0.91(N)  & 1.01(N) & 	   (b)\\
\enddata

{\small
Columns :- (1) source name; (2) date of observation; (3) telescope used; (4) filter used; 
(5) duration of monitoring; (6) number of data points in the DLC; 
(7) rms of the steadiest star-star DLC; (8) INOV amplitude ($\psi$); (9) mean magnitude difference of the t-s1 DLC; 
(10) mean magnitude difference of the t-s2 DLC; (11) mean magnitude difference of the s1-s2 DLC; 
(12) $\chi^2-$ value obtained for the t-s1 DLC (variability status of the DLC); 
(13)  $\chi^2-$ value obtained for the t-s2 DLC (variability status of the DLC);
(14) $\chi^2-$ value obtained for the s1-s2 DLC (variability status of the DLC);
(15) $F-$ value obtained for the t-s1 DLC (variability status of the DLC);
(16) $F-$ value obtained for the t-s2 DLC (variability status of the DLC); 
(17) $F-$ value obtained for the s1-s2 DLC (variability status of the DLC);
(18) modified $C-$ value obtained for the t-s1 DLC (variability status of the DLC);
(19) modified $C-$ value obtained for the t-s2 DLC (variability status of the DLC);
(20) modified $C-$ value obtained for the s1-s2 DLC (variability status of the DLC);
(21) reference for the INOV data.\\
{$^*$} blazars detected at TeV energies. \\
{$^\P$} ST - Sampurnanand Telescope (ARIES); HCT - Himalayan Chandra Telescope (IIA);
IGO - IUCAA Girawali Observatory; VBT - Vainu Bappu Telescope (IIA);  GSO - Gurushikhar telecsope (PRL). \\
$^\dag$ V = Variable; N = Non-variable; PV = Probable Variable; \\
{$^\pounds$}References for the INOV data:
(a) \citet{2005MNRAS.356..607Stalin}; (b) \citet{2007BASI...35..141AG}; 
(c) \citet{2004MNRAS.350..175Stalin}; (d) \citet{2010MNRAS.401.2622AG};
(e) \citet{2004MNRAS.348..176Sagar}; (f) \citet{2012A&A...544A..37AG};
(g) \citet{2011MNRAS.416..101GK}; (h) AGs unpublished data; 
(i) \citet{2004JApA...25....1Stalin}; (j) \citet{2009MNRAS.399.1622AG}.
}
\end{deluxetable}
\end{landscape}
\clearpage
\newpage

\begin{table}
\centering
\caption{Estimates of INOV duty cycle (DC) for our sets of RQQs, RIQs, LDQs, LPCDQs, HPCDQs and TeV detected blazars$^*$.}
\label{dc_stat}
\begin{tabular}{cccc|ccc|ccc}\\\hline
                                &\multicolumn{3}{|c|}{$\chi^2-test$}      &\multicolumn{3}{|c|}{ $F-test$} &\multicolumn{3}{|c|}{ modified $C-test$} \\ \hline
                                & INOV DC     &  INOV DC    & INOV DC    & INOV DC     &  INOV DC    &  INOV DC    & INOV DC     &  INOV DC    &INOV DC   \\
                                & (using s1)  &  (using s2) & (average)  & (using s1)  &  (using s2) &  (average)  & (using s1)  &  (using s2) &(average)  \\
                                &  (per cent) & (per cent)  & (per cent) &  (per cent) & (per cent)  &  (per cent) &  (per cent) & (per cent)  &(per cent)\\
(1)                             &   (2)       & (3)         & (4)        &  (5)        &  (6)        &   (7)       &  (8)        &  (9)        &  (10) \\
\hline                                                                                                                             
{\bf RQQs (68 DLCs)}            &             &             &            &             &             &          &             &             &         \\
for all values of $\psi$ :      & 23(27)      & 21(24)      &  22(25)    & 10(23)      & 11(23)      & 10(23)   & 10(23)      & 11(23)      & 10(23)    \\
\hline
\end{tabular}

$^*$Values inside parentheses are when `PV' cases are also treated as `V' cases. \\
$^\dag$ See Sect. \ref{summary}.\\
Columns : (1) target AGN class; 
(2) INOV DC obtained using the t-s1 AGN DLC, estimated from the $\chi^2-$test; 
(3) INOV DC obtained using the t-s2 AGN DLC, estimated from the $\chi^2-$test; 
(4) average of columns 2 and 3; 
(5) INOV DC obtained using the t-s1 AGN DLC, estimated from the $F-$test; 
(6) INOV DC obtained using the t-s2 AGN DLC, estimated from the $F-$test; 
(7) average of columns 5 and 6; 
(8) INOV DC obtained using the t-s1 AGN DLC, estimated from the modified $C-$test; 
(9) INOV DC obtained using the t-s2 AGN DLC, estimated from the modified $C-$test; 
(10) average of columns 9 and 10. 
 
\end{table}

\clearpage

\section*{Acknowledgements}
AG and G-K would like to thank Dr.\ Santosh Joshi (ARIES) for carrying out 
a few of the optical observations used in the present work. 
The authors are thankful to the anonymous referee for helpful
suggestions.


\end{document}